\documentstyle[preprint,aps]{revtex}

\def\pmb#1{\setbox0=\hbox{$#1$}%
\kern-.025em\copy0\kern-wd0
\kern.05em\copy0\kern-\wd0
\kern-.025em\raise.0433em\box0 }
\def\ttt#1{%
\setbox5=\hbox{$#1$}%
\setbox6=\hbox{\the\scriptfont2\char'030}%
\ifnum\wd5>\wd6{\vbox{\offinterlineskip
\hbox to\wd5{\hfil\the\scriptfont2\char'030\hfil}%
\hbox to\wd5{\hfil\the\scriptfont2\char'030\hfil}%
\vskip1.4pt\hbox{$#1$}}}%
\else{\vtop{\offinterlineskip \hbox{\the\scriptfont2\char'030}
    \hbox{\the\scriptfont2\char'030} \vskip1.4pt\hbox
    to\wd6{\hfil$#1$\hfil}}}\fi}

\newcommand{\gwig}{\mbox{\,\raisebox{.3ex}
{$>$}$\!\!\!\!\!$\raisebox{-.9ex}{$\sim$}}\,}

\def\square{\kern1pt\vbox{\hrule height1.2pt\hbox{\vrule
      width1.2pt\hskip3pt \vbox{\vskip 6pt}\hskip3pt\vrule
      width0.6pt}\hrule height0.6pt}\kern1pt}

\begin{document}

\def\footnoterule{\hrule width \hsize}
\skip\footins = 20pt
\footskip     = 20pt
\footnotesep  = 20pt
\textwidth=6.75in
\hsize=6.9in
\oddsidemargin=0in
\evensidemargin=0in
\hoffset=-.2in

\textheight=9.4in
\vsize=9.4in
\topmargin=-.5in
\voffset=-.3in
\setcounter{page}{0}

\title{ TOY MODELS OF NON-PERTURBATIVE \\
        ASYMPTOTIC FREEDOM IN $\phi_6^3$ }

      \author{J.M. Cornwall\footnote{Electronic address:
          cornwall@physics.ucla.edu} and D.A.
        Morris\footnote{Electronic address: dmorris@physics.ucla.edu} }
      \medskip \address{Physics Department, University of California at
        Los Angeles\\ 405 Hilgard Ave., Los Angeles, CA ~90095--1547}
\maketitle
\begin{abstract}
  We study idealizations of the full nonlinear Schwinger-Dyson
  equations for the asymptotically free theory of $\phi^3$ in six
  dimensions in its meta-stable vacuum.  We begin with the cubic
  nonlinearity and go on to all-order nonlinearities which contain
  instanton effects. In an asymptotically free theory the relevant
  Schwinger-Dyson equations are homogeneous and ultraviolet finite and
  perturbative methods fail from the outset.  We show how our toy
  models of the cubic Schwinger-Dyson equations contain the usual
  diseases of perturbation theory in the massless limit (e.g.,
  factorially-divergent $\beta$-functions, singular Borel-transform
  kernels associated with infrared renormalons) and show how these
  models yield specific mechanisms for removing such singularities when
  there is a mass gap.  The solutions to these homogeneous equations,
  in spite of being ultraviolet finite, still depend on an undetermined
  parameter equivalent to the perturbative renormalization scale $\mu$.
  In the all-order nonlinear equation we show how to recover the usual
  renormalization-group-improved instanton effects and associated
  factorial divergences.
\end{abstract}

\thispagestyle{empty}

\widetext

\vspace{\fill}

\noindent
\hbox to \hsize{UCLA/95/TEP/20
\hfil June 1995}
\vskip-12pt

\newpage
\section{\bf INTRODUCTION}

Asymptotically free theories appear to have perturbatively calculable
properties in the ultraviolet (UV) where the running coupling gets
small.  The well-known price paid for this convenience of perturbation
theory in the UV is its complete breakdown at some low momentum scale,
usually thought of as being of $O(\Lambda_{\rm RG})$ where
$\Lambda_{\rm RG}$ is the renormalization group (RG) mass.  However, as
one goes to higher orders $N$ in perturbation theory, the critical
momentum scale grows exponentially in $N$ essentially because the
number of graphs grows as $N!$.  That is to say, the contribution of
the $N^{\rm th}$ order term in perturbation theory behaves like
$N!(a\bar g^2)^N$ where $\bar g \sim (\ln k^2)^{-1/2}$ is the running
charge; to keep this term small as $N$ grows requires an exponential
increase in $k^2$.  There are also factorial divergence associated with
renormalons\cite{thooft}.  Consequently, perturbation theory to all
orders cannot be used for any momentum scale, however large, in an
asymptotically free theory without understanding how to deal with
non-Borel-summable factorial divergences.  This is currently not a
practical limit on perturbation theory in QCD, where the critical
momentum does not creep into the UV until $N$ is rather large, say $N
\geq 10$.

A more practical issue is understanding QCD processes at infrared (IR)
scales $k \sim \Lambda_{\rm RG}$ and the attendant phenomena of
confinement, condensates, renormalons, large instantons, etc.\cite{zak}
The purpose of the present paper is to discuss the factorial
divergences mentioned above, as well as IR issues (except for
confinement) in a toy model of asymptotic freedom\cite{macfarlane}:
$\phi^3$ in six dimensions $(\phi^3_6)$.  This theory is terminally ill
because the Hamiltonian is unbounded below but we can go quite far
before encountering pathologies.  The idea is to study the theory at
all momentum scales from the IR to the UV in a connected way, without
using the conventional crutches of perturbation theory.  Our study will
use the machinery of nonlinear Schwinger-Dyson equations\cite{cornpap}
for $\phi^3_6$.

Even without the spin complications of QCD, it is too hard for us to
solve the full Schwinger-Dyson equations numerically, even if only the
cubic nonlinearity is saved in the Schwinger-Dyson equation for the
vertex.  So we turn to toy models, first saving only the cubic term for
the vertex and later considering some aspects of the all-order
nonlinearities in the vertex equation.  Although we have no proof, we
believe that these toy models fairly represent qualitatively, but not
quantitatively, all the diseases and their cures we would encounter in
the full Schwinger-Dyson equations.

Of course, any study of $\phi^3_6$ must ultimately break down because
the theory itself is ill-defined; it has no stable vacuum.  We will see
that the breakdown occurs when we try to define the sum of the
non-Borel-summable $N!$ divergences associated with vertex skeleton
graphs having $2N+1$ vertices.  These sums contain the contributions of
instantons\cite{houghton,lipatov}, which we need not add as an explicit
ingredient in analyzing the field theory; they are already in the
Schwinger-Dyson equations.

The Schwinger-Dyson equations of an asymptotically free theory have a
special feature: the renormalization constants which appear in the
renormalized Schwinger-Dyson equations must vanish.  This vanishing is
clear from the canonical form of the $\phi^3_6$ vertex equation (or its
analog in QCD), in which the vertex renormalization constant $Z_1$
appears only as an additive inhomogeneous term.  At large momentum,
asymptotically freedom requires the vertex and all the other terms of
the Schwinger-Dyson equation to vanish, as we will show; this is
inconsistent unless $Z_1=0$.  However, one must be careful.  The
unqualified statement $Z_1=0$ is false and leads to paradoxes.  A more
precise statement resembles that of lattice gauge theory where the
theory is first defined with an UV cutoff $\Lambda_{\rm UV}$ and
$Z_1\rightarrow 0$ at a specific rate as $\Lambda_{\rm UV} \rightarrow
\infty.$

We will see that it is correct simply to set $Z_1=0$ in the canonical
form of the vertex equation (or in an analogous equation for
the running coupling, which is RG-invariant).  The resulting vertex
equation is homogeneous, which completely disconnects it
from any perturbative approach and vastly complicates the analysis
in some ways.

At this point we turn to toy models of the homogeneous and nonlinear
vertex Schwinger-Dyson equation.  Not all models that we have studied
will be presented, since they all show the same features.  We believe
that all these features will occur in the full Schwinger-Dyson equation
as well.  Our studies (both analytic and numeric) fall into several
categories:

1A--{\it Cubic nonlinearity: massless}.  The toy-model vertex Schwinger-Dyson
equation for the running coupling can be converted
to a (nonlinear) ordinary differential
equation.  There are several types of solutions, all of them having
singularities at sufficiently small momenta and having undetermined
parameters.  One of them behaves as expected from perturbation theory
for large momentum where the running coupling varies as
$(\ln k^2)^{-1/2}$; other solutions vanish as
an inverse power of $k^2$ (modulo logarithms).  We find
a differential equation for the $\beta$-function
whose power-series solution diverges
like $N!$, with all terms negative\cite{cornpap};
this divergence is associated with
IR renormalons.  We use the Laplace transform of the vertex equation
in $\ln k^2$ to find an equation for the Borel transform of, e.g., the
$\beta$-function, and find explicitly its pole structure at the IR
renormalon singularity.

1B--{\it Cubic nonlinearity: massive}.  In this case, as is well-known,
all the IR renormalon difficulties are resolved (at least for large
enough mass), and the toy-model Schwinger-Dyson equations are
non-singular.  The only solutions which are regular for all Euclidean
momenta behave like $(\ln k^2)^{-1/2}$ for large $k$.  All solutions
depend on a single continuous real parameter, equivalent to the RG
renormalization point $\mu$; this is somewhat of a surprise, since all
models yield UV-finite solutions needing no cutoff or other
regularization.  Modulo this parameter, the solutions show
uniquely-determined power-law corrections to perturbation theory (e.g.,
condensates and other higher-twist terms).  Generally, the toy models
(as well as the full Schwinger-Dyson equation) cannot be reduced to
differential equations, but there is one notable exception, which we
analyze in some detail.  As for the massless case, this is a
cubically-nonlinear second-order ordinary differential equation.  The
Laplace-transform form of the vertex equation shows that the renormalon
pole singularity is cancelled by another term vanishing like an inverse
power of $k^2$ (modulo logarithms) at large $k$.

2.  {\it All-order nonlinearity}.  First, we prove a theorem (similar
to another used recently\cite{cornmor} to study $N!$ divergences in
$\phi^4_4$) that the imaginary part, in Minkowski space, of every
skeleton graph for the $\phi^3_6$ vertex has neither IR nor UV
logarithms.  The implication is that the UV and IR behavior of the
dressed graphs is entirely determined by the dressed vertex itself,
which allows for an approach to the all-order problem by successive
approximations, beginning by inputting the solutions to the cubic
Schwinger-Dyson equation.  When we use the $(\ln k^2)^{-1/2}$ behavior
found earlier as input, we find that the $N$-vertex graph has the
asymptotic UV behavior $\sim (\ln k^2)^{-(N-2)/2}$.  Using some deep
graph-theoretic results and other tools developed in an earlier
$\phi^4_4$ study, we show that this momentum factor is multiplied by
$a^N N!$, where the positive constant $a$ is rather close to what would
be expected from a one-loop RG-improved instanton analysis.  The sum
over all $N$ is ambiguous, but should have an imaginary part reflecting
the instability of $\phi^3_6$.  The real part shows the usual power-law
behavior expected from instantons.

We comment on some of these results in more detail.  Among our major
results are our demonstrations of i) how the Schwinger-Dyson equation
can be expressed as a nonlinear integral equation for what would be
called the Borel transform in perturbation theory, ii) how the massless
Schwinger-Dyson equation gives rise to poles in this transform and iii)
how the massive Schwinger-Dyson equation gets rid of these poles by
condensate terms.  In its simplest terms, the massless case yields a
Borel integral for the $\beta$-function derived from the running
coupling of the form
\begin{equation}
\label{eq:borel}
\beta (g)={\rm constant}~\times \int^\infty_0 d\alpha
\,\, {H(\alpha)\over 2-\alpha}~e^{-\alpha/(bg^2)}
\end{equation}
where $H(\alpha)$ is regular at $\alpha=2$.  This corresponds
to a $\beta$-function with terms behaving like $-g^{2N+1}(b/2)^NN!$
at large $N$.  Here $b$ is the lowest-order coefficient in the
$\beta$-function:  $\beta=-bg^3+\cdots$.  We show that, when masses
are included, Eq.~\ref{eq:borel} changes to
\begin{equation}
\label{eq:regborel}
\beta (g) ={\rm constant}~\times \int^\infty_0 d\alpha
\, \, {H(\alpha)\over 2-\alpha}
\, \left( e^{-\alpha/(bg^2)}-e^{-2/(bg^2)} \right)
\end{equation}
which has no singularity at $\alpha=2$.  The running coupling has a
similar expression, but with $1/(bg^2)$ replaced by $\ln(k^2+M^2)$;
this shows that the cancellation involves terms vanishing like
$(k^2)^{-2}$ at large $k$.  Of course, the perturbative expansion of
the $\beta$-function is the same as before, but Eq.~\ref{eq:regborel}
shows how that sum is actually defined by the Schwinger-Dyson equation.

Another interesting result is that all our toy models, while perfectly
finite in the UV and requiring no cutoffs or regularizations, still
have a single real parameter in the solutions, which is not fixed by
the vertex Schwinger-Dyson equation above.  This parameter is
equivalent to the usual renormalization-point mass $\mu$ of
perturbation theory, which arises precisely because perturbation theory
requires a cutoff.  We believe that this parameter persists even in the
homogeneous, finite Schwinger-Dyson equation because, as mentioned
above, the categorical statement $Z_1=0$ is not true.  There are other
forms of the vertex Schwinger-Dyson equation in which it is essential
to introduce a cutoff and cancel the cutoff dependence of $Z_1$ against
that of various integrals involving the vertex.  In other words, the
homogeneous finite Schwinger-Dyson equation we use recognizes its
heritage as a renormalizable, not super-renormalizable, theory.  It is,
of course, possible that combining the study of the Schwinger-Dyson
vertex with some other Schwinger-Dyson equation of the theory would fix
the free parameter we find, but we have no evidence for that.

A final result worth commenting on is the partial uncovering of
instanton phenomena in sums of graphs---not just bare graphs, but
dressed graphs, which leads to the incorporation of what would be
called one-loop RG effects at our level of investigation.  It is not
straightforward to deal with instantons in the usual way, improving
semi-classical results with the perturbative RG.  For one thing, these
perturbative corrections to instantons are not under control any more
than they are in other sectors of the theory; there are IR renormalons
which must be tamed by masses as in Eq.~\ref{eq:regborel}. Furthermore,
there are no exact instantons in the massive theory and perhaps most
crucial, standard instanton techniques are really only applicable when
all external momenta vanish\cite{corntik}.  Ultimately, instanton
effects at all momentum scales will have to be dealt with by
Schwinger-Dyson methods.  What we leave completely open here is how
unitarity or some other physical effect provides a definition for the
divergent sums associated with instantons.  One would not be surprised
to find a modification like Eq.~\ref{eq:regborel} entering for
instantons as well as IR renormalons, at least in a well-behaved theory
like QCD.

Which of our results might persist in the theory of real interest to
us, QCD?  Even forgetting about confinement, the major complication
which sets QCD apart from $\phi^3_6$ is gauge invariance.  Using the
pinch technique\cite{cornpap,corn}, however, it is possible to define
gauge-invariant proper vertices and self-energies which are related by
naive (ghost-free) Ward identities.  For these special vertices and
self-energies $Z_1=Z_2$, and both these renormalization constants
vanish like $(\ln \Lambda_{\rm UV}^2)^{-1/2}$ as $\Lambda_{\rm
  UV}\rightarrow\infty$; one can, in principle at least, write down
finite homogeneous Schwinger-Dyson equations for the gauge-invariant
vertex corresponding to the running coupling, and these should show
phenomena similar to what we did for $\phi^3_6$.  For example,
generation of a QCD mass scale is expected to yield a renormalon
cancellation mechanism like that expressed in Eq.~\ref{eq:regborel} for
Borel-transform renormalon poles; indeed, this mechanism has already
been invoked \cite{bogo} on other grounds.  It is also reasonable to
expect that we can find instanton phenomena in graphical sums, just as
for $\phi^3_6$.

As for persistence of a free parameter like $\mu$, the situation is not
so clear.  The vertex and propagator are not independent as they are in
$\phi^3_6$, which might lead to further constraints.  One expects that
QCD (without quarks) has only one free parameter, the RG mass
$\Lambda_{\rm RG}$, in terms of which all physical quantities are
determined.  This does not seem to be the case for $\phi^3_6$, which
has both $\mu$ and the mass $M$ free, at least at the level of our
investigation.

\section{\bf PRELIMINARIES}

In this section we make some general remarks about the nature of the
perturbation series for $g\phi^3$ theories.  We encounter
non-Borel-summable series in $g$ of the usual asymptotically free type,
but in certain cases these series define one or more entire functions
of $g$.  We also set the stage for the investigation of $\phi^3_6$,
discussing some aspects of the Schwinger-Dyson equations for the proper
vertex, the proper self-energy, and a vertex formed from these which is
RG invariant and corresponds to the running charge.

\subsection{General $\phi^3$ Theories}

Theories of $\phi^3$ type are well-defined, in general, only for purely
imaginary coupling where they sometimes make sense physically
\cite{fisher,abarbanel}. It is sometimes possible to continue the
theory to real $g$ by imposing definitions (typically, the order in
which integrals are to be done) which may or may not make sense
physically.  Consider as an example a zero-dimensional model with two
fields $x$ and $y$ with a partition function given by
\begin{equation}
\label{eq:zerodim}
Z(g)=\int^\infty_{-\infty}~dxdy~ e^{-x^2-y^2+gx^2y}
\end{equation}
which is well-defined if $g$ is imaginary.  We can give two forms to
$Z$ depending on which order the integrals are done; integrating
over $y$ first gives
\begin{equation}
\label{eq:intfirst}
Z(g) = \sqrt{\pi}~\int^\infty_{-\infty}~dx~ e^{-x^2+g^2x^4/4}
\end{equation}
which is a wrong-sign $\phi^4$ theory.  Expanding Eq.~\ref{eq:intfirst}
in powers of $g^2$ gives rise to a typical asymptotically free series
with terms behaving like $N!~N^b~(a g^2)^N$ where $a$ and $b$ are fixed
constants with $a>0$.  It is not obvious how such a divergent series is
to be defined.

If we instead perform the integral over $x$ first in
Eq.~\ref{eq:intfirst} we get
\begin{equation}
\label{eq:intsecond}
Z(g) = \sqrt{\pi}~\int^\infty_{-\infty}~dy(1-gy)^{-1/2}~e^{-y^2}~.
\end{equation}
This appears to be singular at $y=1/g$ (the classical saddle point) of
Eq.~\ref{eq:zerodim} but in fact $Z(g)$ as defined by
Eq.~\ref{eq:intsecond} is an entire function of $g$: as $g$ changes
from a pure imaginary value to the real axis, or anywhere else in the
complex $g$-plane, the contour can be deformed to avoid the potential
singularity.  Actually, Eq.~\ref{eq:intsecond} defines two entire
functions, depending on whether one begins with $g$ on the positive or
negative imaginary axis.  One or the other of these possibilities may
be singled out by physical considerations, or a linear combination may
be used (e.g., the average of the two entire functions is real for real
$g$).

These remarks can be generalized to $d=6$ theories of the type
$g\phi|\psi|^2$ where $\phi$ is a single-component Hermitian field, and
$\psi$ is a single-component complex field.  The Euclidean partition
function is
\begin{equation}
Z = \int (d\phi d\bar\psi d\psi)
\exp
\left[  -\int d^6x
\left( |\partial_\mu\psi|^2
+{1 \over 2}(\partial_\mu \phi)^2+g\phi|\psi|^2
+{\rm mass~terms} \right)
\right]~.
\end{equation}
The functional integral over $\phi$ is free, and when done reveals a
wrong-sign non-local $|\psi|^4$ theory for $\psi$, analogous to
Eq.~\ref{eq:intsecond}.  But doing the (free-field) functional integral
over $\psi$ gives
\begin{equation}
Z\sim \int(d\phi)~\exp \left[ -\int d^6x
\left( {1\over 2}
(\partial_\mu \phi)^2+{1\over 2}M^2\phi^2-{\rm Tr}~\ln
(-\square+m^2+g\phi)
\right)
\right]~.
\end{equation}
Superficially this would appear to define an entire function of $g$,
since the quadratic terms in the action dominate the logarithm.
However, the argument of the logarithm may vanish (when $\phi\sim
M^2/g$) and no firm conclusion can be drawn.  Moreover, dimensional
transmutation tells us that $g$ must disappear, to be replaced by a
running coupling.  Still, it is intriguing to speculate that
asymptotically free theories in general may have matrix elements and
Green's functions which are entire functions of the running charge.  If
some such speculation is correct, it can only be developed by methods
completely divorced from perturbation theory.

\subsection{The Schwinger-Dyson Equations for $\phi^3_6$}

Throughout this paper (except for section IV)
we work only at Euclidean momenta; the action
for our theory is
\begin{equation}
S = \int d^6x\biggl[{1\over 2}(\partial_\mu\phi)^2+
   {M^2\over 2}\phi^2+{g\over 3!} \phi^6\biggr]~.
\end{equation}
We initially focus our attention on the Schwinger-Dyson equations for
the renormalized propagator $\Delta$ and proper vertex $\Gamma$.  Later
we will construct the running coupling from a particular combination of
these quantities. The equation for $\Gamma$ can be written in several
equivalent ways. In Fig.~1 we express $\Gamma$ in terms of a four-point
function which is two-particle-irreducible in the $k_1$ channel.  The
four-point function can then be decomposed in terms of skeleton graphs
and three-point functions.  In this and other figures, a small blob
denotes $\Gamma$, a line denotes $\Delta$ (both of which are
renormalized quantities) and $Z_1$ is the vertex renormalization
constant. Figure~2 shows the analogous equations for $\Gamma$ in terms
of the one-particle-irreducible four-point function and Fig.~3 shows
the equation for the inverse propagator.

One of our major concerns will be that $Z_1$ vanishes in an
asymptotically free theory.  This is not entirely an elementary matter
since uncritically setting $Z_1=0$ in Fig.~2 or the first equation for
$\Delta^{-1}$ in Fig.~3 appears to give zero. The resolution, of
course, is that the theory is defined first with an ultraviolet cutoff
$\Lambda_{\rm UV}$ which renders $Z_1$ finite.  In reality, as
$\Lambda_{\rm UV}\rightarrow\infty$ the product of $Z_1$ with a
quantity involving a single bare vertex actually remains finite since
the $\Lambda_{\rm UV}$ dependence of such quantities cancels that of
$Z_1.$ It is permitted, however, to set $Z_1=0$ in those forms of the
Schwinger-Dyson equations in which no bare vertex appears (e.g., for
$\Gamma$ in Fig.~1 and the second form for $\Delta^{-1}$ in Fig.~3).
The reason is that in an asymptotically free theory the remaining
dressed graphs of the Schwinger-Dyson equation are all individually
finite.  This does not mean that the Schwinger-Dyson equations are
unambiguously finite, since the Schwinger-Dyson equation for $\Gamma,$
expressed in terms of $\Gamma$ and $\Delta,$ has infinitely many terms
and the sum actually diverges, as we discuss in Section~IV.

Let us concentrate for the moment on the lowest-order Schwinger-Dyson
equations for $\Gamma$ and $\Delta^{-1}$, shown in Fig.~4, where we
have now set $Z_1=0$.  Introduce the notation
\begin{equation}
\Delta^{-1}(k^2)=(k^2+M^2) Z_2(k^2)
\end{equation}
where $M^2$ is the renormalized mass.  We anticipate, guided by
one-loop RG-improved perturbation theory\cite{macfarlane}, that the
solutions to the Schwinger-Dyson equations of Fig.~4 behave in the UV
($k^2\gg M^2$) as various powers of a logarithmic function $D(k^2)$:
\begin{eqnarray}
\label{eq:ddef}
D(k^2)  =  1 + bg^2 \ln(k^2/M^2) \\
\Gamma \rightarrow D^{-\alpha},     \qquad
Z_2 \rightarrow D^{-\gamma}~.
\end{eqnarray}
In Eq.~\ref{eq:ddef}, $b$ is a positive number given below; it is the
same coefficient as in the one-loop $\beta$-function
$\beta=-bg^3+\cdots$.  Any finite mass may be used to set the argument
of the logarithm in Eq.~\ref{eq:ddef} and in consequence the 1 on the
right hand side of this equation is irrelevant; we will drop it.

Insert these forms into the equation for $\Gamma$ of Fig.~4a
using a simplified form suitable for finding the UV behavior:
\begin{equation}
\label{eq:gamdef}
\Gamma(k^2)={\pi^3g^2 \over 2(2\pi)^6} \int^\infty_{k^2}
{dp^2\over p^2}(bg^2 \ln p^2)^{3\gamma-3\alpha}=
(bg^2 \ln k^2)^{-\alpha}~.
\end{equation}
Doing the integral reveals consistency, provided that
\begin{eqnarray}
\label{eq:consist}
1-2\alpha+3\gamma&=&0 ,\\
2(4\pi)^3 \alpha b & =& 1~.
\end{eqnarray}
Note that if the condition of Eq.~\ref{eq:consist} is satisfied, $g^2$
drops out of the equation so we might as well drop it in $D$ of
Eq.~\ref{eq:ddef}, along with the 1 previously dropped.  That is, there
really is no coupling constant in an asymptotically free theory, only
mass scales.

We need one more equation to determine $\alpha,\gamma$, and $b$, which
is furnished by the $\Delta^{-1}$ equation.  This is slightly ticklish
because it has a quadratic divergence, subsumed by mass
renormalization.  In a $g\phi|\psi|^2$ theory one may avoid direct
consideration of the $\Delta^{-1}$ equation by coupling $\psi$ to the
electromagnetic field in the usual way, computing the lowest-order
electromagnetic vertex shown in Fig.~5, using the Ward identity
\begin{equation}
q^\mu\Gamma^{(\gamma)}_\mu(q,p,p^\prime)=
\Delta^{-1}(p)-\Delta^{-1}(p^\prime)
\end{equation}
and then setting $p^{\prime^2}=M^2$, so $\Delta^{-1}(p^{\prime^2})=0$.
It turns out (see below) that $Z_2=0$ also, so the $\Gamma^{(\gamma)}$
equation is homogeneous of Baker-Johnson-Willey type.  We leave it to
the reader to conduct an analysis like that which led to
Eq.~\ref{eq:gamdef}; it gives the same result as Eq.~\ref{eq:new}
below.

Either as sketched above or by analysis of the $\Delta^{-1}$ equation
in Fig.~3 one finds a new condition
\begin{equation}
\label{eq:new}
(4\pi)^3 \gamma b=1/12~.
\end{equation}
Then one finds
\begin{equation}
\alpha=2/3~, \quad \gamma=1/9~, \quad
b={3\over 4(4\pi)^3}
\end{equation}
which are the usual results\cite{macfarlane} of one-loop RG-improved
perturbation theory, with $\beta$-function $-bg^3+\cdots$.  These
constants also appear in the cutoff dependence of the perturbative
renormalization constants
\begin{equation}
Z_1  = \left( 1+bg^2 \ln(\Lambda_{\rm UV}^2/\mu^2) \right)^{-2/3},
\qquad\qquad
Z_2  = \left( 1+bg^2 \ln(\Lambda_{\rm UV}^2/\mu^2) \right)^{-1/9}
\end{equation}
both of which vanish as the cutoff $\Lambda_{\rm UV} \rightarrow \infty$.

At this point it is convenient to introduce a special combination of
$\Gamma$ and $Z_2$, corresponding to the running charge
\begin{equation}
\label{eq:running}
\bar g(k_1,k_2,k_3)\equiv g \,
\displaystyle{ \Gamma(k_1,k_2,k_3) \over
\sqrt{ Z_2(k_1) Z_2(k_2) Z_2(k_3) }   }~.
\end{equation}
We keep the explicit $g$ only to observe that, with $D\equiv bg^2\ln
k^2,~\Gamma \sim D^{-2/3},~Z_2\sim D^{-1/9}$, all the $g$'s cancel in
Eq.~\ref{eq:running}.  We now find the asymptotic behavior, when all
$k_i$ scale like a large momentum $k$,
\begin{equation}
\bar g^2 \rightarrow \displaystyle{1\over b \ln k^2 }~.
\end{equation}
This is the usual one-loop running coupling.  If constructed in
perturbation theory, $\bar g$ would be explicitly independent of the
renormalization point $\mu$.

With $Z_1=0$, the Schwinger-Dyson equation for $\bar g$ (the solid blob
in Fig. 6) looks like that for $\Gamma$ with the important difference
that now the propagator lines correspond to free propagators with the
renormalized mass.  This seems to lead to a quite remarkable
circumstance in which the Schwinger-Dyson equation for $\bar g$ is
completely self-contained.  However, this is misleading since the
infinite series of terms in Fig.~6 does not have a sum which is
well-defined.  In fact, this series contains the basic instanton
phenomenon of $\phi^3_6$, as we will see in Section IV, but we will
never need to invoke instantons explicitly.  We will also see in
Section~IV that Schwinger-Dyson graphs for $\bar g$ having $2N+1$
insertions of $\bar g$ behave at large momentum like $(\ln
k^2)^{-N/2}$, so that the leading UV behavior is correctly captured in
the cubic graph.

In the next section we will investigate toy models of the
Schwinger-Dyson equation somewhat similar to the integral of
Eq.~\ref{eq:gamdef}.  If we uncritically write such an equation for
$\bar g$ with free propagators $(\gamma=0)$ and test whether $\bar
g=(b\ln k^2)^{-1/2}$ is a solution, we find that everything works
except that the integral is 4/3 larger than it should be.  This is
because the simple form assumed for $\bar g$ does not distinguish the
momenta on the legs of $\bar g$ which are in general different; this
distinction matters for $\bar g$ with its separate powers of
$Z_2(k_i)^{-1/2}$, not all of which involve momenta which are
integrated over.  In the toy models of the next section we will
compensate for this by multiplying the right hand side of integrals
cubic in $\bar g$ by 3/4.

Why do we need toy models at all?  The answer is that even without the
spin complications of QCD, and saving only cubic nonlinearities, the
full Schwinger-Dyson equation for $\phi^3_6$ is quite complicated.
Here we set up this cubic equation schematically, before going on to
the toy models in the next section.

The proper vertex $\Gamma(k_1,k_2,k_3)$ depends on three scalar
variables $k_1^2,k_2^2,k_3^2$ in a way which is constrained by
causality and spectrum conditions.  These may be enforced with the
Nakanishi representation\cite{nakanishi}, which is derived from the
Feynman-parameter representation of graphs but which is expected to
hold even for non-perturbative processes.  For the vertex it reads:
\begin{equation}
\label{eq:gamdef2}
\Gamma(k_1,k_2,k_3)=\int dz_1 dz_2 dz_3 d\sigma
\delta( 1 - z_1 - z_2 - z_3 )
{h(\sigma,z_1,z_2,z_3)\over
\sigma+ z_1 k_1^2 + z_2 k_2^2 + z_3 k_3^2}~.
\end{equation}
The spectrum is expressed through the support of $h$.  In a massive
theory, the lower limit of the $\sigma$ integral depends on masses and
$z_i$ whereas the upper limit is infinity.  The asymptotic behavior
$\Gamma\sim (\ln k^2)^{-2/3}$ requires that $h\sim (\ln \sigma)^{-5/3}$
at large $\sigma$ (in perturbation theory $h\rightarrow {\rm
  constant}$, and Eq.~\ref{eq:gamdef2} needs the usual subtraction).
One then uses Eq.~\ref{eq:gamdef2} and the Lehmann representation for
the propagator to write out the Schwinger-Dyson equation of Fig.~6.  If
the loop momentum is called $p$, the momentum integral depends on three
scalars $(p^2,p\cdot k_1,p\cdot k_2)$.  One might then try to solve the
equations numerically, but it is a substantial project. If each $k_i^2$
takes on 100 values, one has to do $10^6$ three-dimensional integrals
numerically for every iteration.  We have not attempted this, but have
studied some toy models which we are convinced have the qualitative
features as the real (cubic) Schwinger-Dyson equation.  We now turn to
those models.

\section{\bf THE TOY MODELS}

We have examined several models based on simplifying the cubic
Schwinger-Dyson equation of Fig. 1; we will describe a few here.  A
simplification common to all the models is to assume that
$\Gamma(k_1,k_2,k_3)$ depends only on $k_1^2+k_2^2+k_3^2$ which
corresponds to fixing all $z_i=1/3$ in the Nakanishi representation of
Eq.~\ref{eq:gamdef2}.

In this section we restrict ourselves to the special momentum
configuration $k_1=0,k_2=-k_3=k$, and we will be interested in the
whole (Euclidean) range $0\leq k^2 \leq\infty$.  We will always use
free propagators in the model Schwinger-Dyson equation, which can
either be interpreted as an approximation to the propagators in the
Schwinger-Dyson equation for the proper vertex $\Gamma$, or as a
special kind of simplification of the Schwinger-Dyson equation for the
running coupling $\bar g$, where free propagators are required (see the
discussion around Eq. \ref{eq:running}), but where the assumption of
dependence of $\bar g$ only on $\sum k_i^2$ cannot really be correct
because of the $(Z_2(k_1) Z_2(k_2) Z_2(k_3))^{-1/2}$ factor in its
definition in Eq.~\ref{eq:running}.  We choose to present our results
as referring to $\bar g$; the error committed in assuming $\bar g$
depends only on $\sum k_i^2$ is quantitative but does not affect our
major qualitative conclusions.  The reason for our choice of
interpretation is that all model solutions behave like $\bar g$ in the
UV region, that is, like $(\ln k^2)^{-1/2}$.

Various models differ from each other by further simplification of the
momenta appearing in internal vertices.  The Schwinger-Dyson equation
with correct momentum assignments is shown in Fig.~7, and corresponds
to the integral equation
\begin{equation}
\label{eq:twodim}
G(2k^2)= \int \displaystyle{ d^6p \over \pi^3} \,\,
\displaystyle{ G(2(p+k)^2)
G^2 \left( p^2+k^2+(p+k)^2 \right)
\over \left( (p+k)^2 + m^2 \right)^2 (p^2+M^2) } ~.
\end{equation}
where the normalization is chosen so that $G\rightarrow (\ln
k^2)^{-1/2}$ for $k^2 \gg M^2,m^2$ so that we interpret $G$ as
$b^{+1/2}\bar g$.  Later we will present numerical solutions of
Eq.~\ref{eq:twodim}, but for now it is much more important to have an
analytic understanding of simplified versions of this equation. The
reason for using two masses $M$ and $m$ will be made clear later.

\subsection{\bf The Massless Model.}

The first model to be considered is the massless model ($M=m=0$) where,
to make things more tractable, we set $k=0$ in the arguments of $G$ on
the right hand side of Eq.~\ref{eq:twodim} to get
\begin{equation}
\label{eq:onedim}
G(2 k^2) =
\int {d^6p \over \pi^3 } \displaystyle{ G^3(2 p^2) \over
(p+k)^4 p^2 }~.
\end{equation}
It can be shown, by an uninteresting argument bounding $G$ from above
and below, that the solutions to Eq.~\ref{eq:twodim} and
Eq.~\ref{eq:onedim} have the same behavior for large $k$.

Performing the angular integrals in Eq.~\ref{eq:onedim} and introducing
the variables
\begin{equation}
\label{eq:tdef0}
t  = \ln( k^2 / \mu^2 )\qquad
t'  = \ln( p^2 / \mu^2 )
\end{equation}
where $\mu$ is an arbitrary
mass scale, we obtain the one-dimensional integral equation
\begin{equation}
G(t) = {1 \over 2} \int_{-\infty}^{\infty} dt' \, G^3(t')
     + {1 \over 2 } \int_{-\infty}^t dt' \, e^{2(t-t')} G^3(t') .
\end{equation}
Differentiating twice, one finds that $G$ obeys the equation
\begin{equation}
\label{eq:diffeq}
\ddot G+2\dot G  = -G^3
\end{equation}
where $\dot G\equiv dG/dt$. This same equation could be obtained by
noting in Eq.~\ref{eq:onedim} that $\bigl[-4\pi^3(p+k)^4\bigr]^{-1}$ is
the inverse of the d'Alembertian operator in six dimensions.

We must point out that Eq.~\ref{eq:diffeq} actually makes no sense if
we require $G(t) \ge 0 $ (as physical considerations suggest).  Perhaps
the most intuitive way to see this is to think of Eq.~\ref{eq:diffeq}
as the differential equation governing the `position' $G$ of an
anharmonic oscillator as a function of `time' $t$: except for the
trivial solution $(G=0)$ all solutions to this equation inevitably
develop unbounded oscillations as $t\rightarrow -\infty.$ The
problematic region $t \rightarrow -\infty$ corresponds to an IR
divergence at $k=0$. It is this divergence which is responsible for the
renormalon behavior we will shortly encounter.  The $t \rightarrow
+\infty$ solutions are candidates for the large-$k$ behavior for the
massive equation of Eq.~\ref{eq:twodim}.

There are at least two fundamental types of large-$t$ solutions to
Eq.~\ref{eq:diffeq}, and another which is a hybrid of the first two.
In the first type, the $\ddot G$ term in Eq.~\ref{eq:diffeq} is treated
as a perturbation to the equation $2 \dot G=-G^3$ which has a solution
$G = t^{-1/2} =(\ln(k^2/\mu^2))^{-1/2}$, as anticipated from
field-theoretic perturbation theory.  Including the $\ddot G$ term as a
perturbation, we exhibit several more terms of this solution,
\begin{equation}
\label{eq:sol1}
G=D^{-1/2}~; \quad
D=t+{3\over 4}~\ln c_1t-{15\over 8t}
\biggl(1-{3\over 10} \ln c_2t\biggr)+\ldots~.
\end{equation}
Here the $c_i$ are constants not determined by consideration of large
$t$ alone.

The behavior of Eq.~\ref{eq:sol1} is, except for the numerical
coefficients, precisely what the RG and perturbation theory would give
for the running coupling.  In fact, it is possible to find the
$\beta$-function for the vertex $\bar g$ by noting that
\begin{equation}
\beta(\bar g)=2\bar g=2b^{-1/2}\dot G
\end{equation}
which leads immediately from Eq.~\ref{eq:diffeq} to a first-order
differential equation for $\beta(g)$:
\begin{equation}
\beta\biggl(1+{1\over 4}{d\beta \over dg}\biggr)=-bg^3~.
\end{equation}
The power series solution beginning with $bg^3$ shows a
non-Borel-summable factorial divergence associated with renormalons,
\begin{equation}
\label{eq:nbsr}
\beta(g) \dot = -\sum N!~\biggl({b\over 2}\biggr)^N
g^{2N+1}~.
\end{equation}
where, for large $N$, $\dot =$ means equality up to fixed powers of $N$
and overall multiplicative constants.  We will return to this solution
and the associated $\bar g$ shortly, showing how to set up
Eq.~\ref{eq:diffeq} directly as an integral equation for a Borel
transform.

There is another class of solutions to Eq.~\ref{eq:diffeq} which shows
no relation to perturbation theory; this class arises from by treating
the $-G^3$ term in Eq.~\ref{eq:diffeq} as a perturbation. The solution
reads
\begin{equation}
\label{eq:sol2}
G = \displaystyle{ A~e^{-2t}\over D_1};
\qquad
D_1 = 1+\displaystyle{A^2\over 24}~e^{-4t}+
\displaystyle{A^4\over 5760}~
e^{-8t}+\displaystyle{A^6~e^{-12t}\over 967680} + \ldots~.
\end{equation}
where $A$ is an arbitrary positive constant.  Note that this solution
behaves as inverse powers of $k^4$, powers which one would like to
associate with condensates and higher-twist terms.

Finally, there is a hybrid solution combining the previous types
of solutions of Eq.~\ref{eq:sol1} and Eq.~\ref{eq:sol2},
\begin{equation}
G=D_2^{-1/2}; \quad
D_2=t+{3\over 4} \ln c_1t+\ldots+2At^3~e^{-2t}+\ldots~,
\end{equation}
where again $c_1$ and $A$ are arbitrary.  The solutions of the real
Schwinger-Dyson equation also ought to have a combination of
perturbative powers of logarithms and powers of $k^2$.  The existence
of such hybrid solutions show that more numerical analysis is not very
revealing, since it is hard to separate powers and exponentials of $t$
(i.e., of $\ln k^2$).

An interesting way of looking at our models is to introduce the Laplace
transform of $G(t)$:
\begin{equation}
\label{eq:lapt}
G(t)\equiv \int^\infty_0~d\alpha~F(\alpha)~e^{-\alpha t}~.
\end{equation}
Since in the lowest order of approximation $t \sim (b\bar g^2)^{-1}$,
the Laplace transform $F(\alpha)$ can also be interpreted as the Borel
transform.  At this order this appears to be a trivial matter, since
$G=b^{1/2}\bar g,~t=(b\bar g^2)^{-1}$ leads to
$F(\alpha)=(\pi\alpha)^{-1/2}$.  But as one goes on more interesting
phenomena appear.

By recognizing $e^{-\alpha t}=(k^2/\mu^2)^{-\alpha}$ and using standard
integration formulas, one can express the massless integral equation
Eq.~\ref{eq:onedim} in the form
\begin{equation}
\label{eq:lapint}
F(\alpha)  =   \displaystyle{\alpha\over 2-\alpha} \int
[dy]~F(\alpha y_1)  F(\alpha y_2)  F(\alpha y_3)
\end{equation}
where $ {[dy]} \equiv \int^1_0~dy_1 \, dy_2 \, dy_3~\delta(1-\sum
y_i).$ Here $y_i$ are Feynman parameters used to express the Laplace
transform variables $(\alpha_i=\alpha y_i)$ of the three vertices on
the right hand side of Eq.~\ref{eq:onedim}.  Perturbation theory
corresponds to $\alpha \ll 1$, in which instance we replace $2-\alpha$
by 2, and verify that $F=(\pi\alpha)^{-1/2}~({\rm i.e.},
G=t^{-1/2})$ is the
small-$\alpha$ solution.  But there clearly is some sort of singularity
at $\alpha=2$, which actually causes Eq.~\ref{eq:twodim} to be
meaningless (just as integrals leading to IR renormalons are
meaningless).  Consider the ``approximation" of replacing the $F(\alpha
y_i)$ by their perturbative values on the right hand side of
Eq.~\ref{eq:lapint}; the result is an $F(\alpha)$ behaving like
$\alpha^{-1/2}/(2-\alpha)$.  We can then construct the Laplace
transform for $\beta(\bar g) \sim \dot G$, and in $e^{-\alpha t}$
replace $t$ by $(b\bar g^2)^{-1}$.  The result is a Borel-transform
representation of $\beta(g)$,
\begin{equation}
\beta(g) \sim \int^\infty_0~d\alpha  \, {\alpha^{1/2}\over 2-\alpha}~
e^{-\alpha/bg^2}~.
\end{equation}
The pole at $\alpha=2$ is, of course, what gives rise to the
factorially-divergent behavior of $\beta$ shown in Eq.~\ref{eq:nbsr}.

A major theme of our work is the demonstration of how such
Borel-transform singularities are cured in the massive theory, as we
know they must be.  So now we turn to massive models.

\subsection{\bf Massive Models.}

We now return to the massive equation Eq.~\ref{eq:twodim}.  If $m\not=
0$ it is no longer possible to find a local differential equation as we
did for the massless case, and analytic progress beyond the leading UV
behavior seems impossible.  However, when $m=0$ we can convert
Eq.~\ref{eq:twodim} to a differential equation (or a corresponding
integral equation in one variable) just as we did when both $M$ and $m$
were set to zero.  As long as $M\not= 0$, there is no IR singularity in
Eq.~\ref{eq:twodim}, as one easily checks by looking at $k=0$.  We
will, using the Laplace transform technique, translate this lack of IR
divergence into a cancellation mechanism for the poles of the
Laplace-transform kernel which appeared in the massless case (see
Eq.~\ref{eq:lapint}).

We have actually looked at several variants of Eq.~\ref{eq:twodim},
which differ in the choice of $m$ and of the arguments of the $G$'s
under the integral.  These are conveniently summarized via:
\begin{equation}
\label{eq:variants}
G(2 k^2)= \int
\displaystyle{ d^6p \over \pi^3 }
\displaystyle{ F(p,k)\over
\left( (p+k)^2+m^2 \right)^2 (p^2+M^2) } .
\end{equation}
The models we will discuss are, for $M\ne0:$
\begin{description}
\item A) $m=0,~F=G^3(2 p^2)~.$
\item B) $m=M,~F=G^3(2 p^2)~.$
\item C) $m=M,~F=G\left(2(p+k)^2 \right)~G^2(p^2+k^2+(p+k)^2))$.
\end{description}
The $k-$independence of $F$ in models A and B gives rise to
one-dimensional integral equations which we analyze both analytically
and numerically. Model C, on the other hand, is Eq.~\ref{eq:twodim}
with $m=M$ which gives rise to a two-dimensional integral equation
which we attack numerically. In addition to models A-C we have also
examined intermediate cases where, for example, $F=G^3\left(2(p+k)^2
\right)$ with $m=M$ or $F=G^3(2(p^2+k^2))$ with $m=M.$ These
additional cases introduce a $k-$dependence into $F$ which is still
mild enough to yield one-dimensional integral equations. However, since
these additional models have all the same qualitative features as
models A-C but yield little extra insight, we will not discuss them
further.

Only model A will be discussed in much detail.  One can show that all
models lead to the same leading UV behavior ($G\rightarrow (\ln
k^2)^{-1/2}$), and numerical work displayed later shows that the models
differ quantitatively but not qualitatively in the IR.  All have the
curious feature, mentioned in Section~I, that the solutions to each
model depend on a single real parameter.

Consider now model A.  Similar to our treatment of the massless model,
we introduce the variables
\begin{equation}
\label{eq:tdef}
t = \ln \biggl(\displaystyle{k^2+M^2\over M^2}\biggr),  \qquad
t^\prime = \ln \biggl(\displaystyle{p^2+M^2\over M^2}\biggr)
\end{equation}
where we have fixed the analog of the arbitrary mass scale $\mu$ in
Eq.~\ref{eq:tdef0} to be equal to $M.$
In terms of these variables, Eq.~\ref{eq:variants} for model A reads
\begin{equation}
\label{eq:diffsep2}
G(t)={1\over 2}\int^\infty_0 dt^\prime~G^3(t^\prime)+{1\over 2}
\int^t_0 dt^\prime\biggl[\biggl({e^{t^\prime}-1\over
e^t-1}\biggr)^2-1\biggr] G^3(t^\prime)~.
\end{equation}
which is equivalent to the differential equation
\begin{equation}
\label{eq:secdiff}
(1-e^{-t})\ddot G  +
(2+e^{-t})\dot G
=-G^3
\end{equation}
with the boundary conditions
\begin{equation}
\label{eq:bc}
G(0) = \displaystyle{1 \over 2}  \int_0^\infty dt' G^3(t'), \qquad
G(\infty)=0~.
\end{equation}
This differential equation differs from the massless equation
Eq.~\ref{eq:diffeq} by the appearance of $e^{-t}$ in the
coefficients; it can, of course, be derived directly from
Eq.~\ref{eq:diffsep2}, just as we did for the massless
equation Eq.~\ref{eq:diffeq}.

It is not hard to see that $G(\infty)=0$ is not really a boundary
condition since all solutions of Eq.~\ref{eq:secdiff} vanish at
$t=\infty$.  Moreover, Eq.~\ref{eq:bc} does not constrain $G(0)$; as
long as $G(0)>0$, there is a solution to Eq.~\ref{eq:secdiff}
satisfying Eq.~\ref{eq:bc}.  So $G(0)$ itself can be chosen as the
parameter of the solution. Once $G(0)$ is picked, the solution is
unique.  All of the models listed above have this feature.

To see how this parameter arises for Eq.~\ref{eq:diffsep2}, consider a
slightly different equation with an explicit parameter $a$:
\begin{equation}
\label{eq:aeq}
\tilde G(t,a)=a-{1\over 2}\int^t_0 dt^\prime \tilde G(t^\prime,a)^3
+{1\over 2}\int^t_0 dt^\prime\biggl({e^{t^\prime}-1\over
  e^t-1}\biggr)^2 \tilde G(t^\prime,a)^3~.
\end{equation}
For $t\gg 1$ the factor $(e^{t^\prime}-1)^2/(e^t-1)^2$ is exponentially
peaked at $t^\prime=t$ so it is a good approximation to evaluate the
relatively slowly varying $\tilde G(t^\prime,a)^3$ at $t^\prime=t$ and
pull it outside the integral to get
\begin{equation}
\label{eq:aeq2}
\tilde G(t\gg 1,a)=a-{1\over 2} \int^{t\gg 1}_0 dt^\prime
G(t^\prime,a)^3+{1\over 8}\tilde G(t\gg 1,a)^3~.
\end{equation}
If we use the fact that $\tilde G(t\gg1,a) \sim 1\sqrt{t}$ and let $t$
go to infinity, we find
\begin{equation}
  a={1\over 2}\int^\infty_0 dt^\prime \tilde G(t^\prime,a)^3~.
\end{equation}
In other words, if $\tilde G(t,a)$ is a solution to Eq.~\ref{eq:aeq}
then $G(t,a)$ automatically satisfies Eq.~\ref{eq:diffsep2}.

We do not know how to solve Eq.~\ref{eq:secdiff} analytically, so yet
another presentation of its contents is needed to understand the taming
of IR renormalons.  We proceed as in the massless case by introducing
the Laplace transform Eq.~\ref{eq:lapt}, with $t$ given by
Eq.~\ref{eq:tdef}.  Introducing a Feynman parameter $x$ for the
$p^2+M^2$ propagator and performing the momentum-space integral, one
gets
\begin{equation}
  G(t)=\int^\infty_0 d\alpha\, \alpha \int [dy] F(\alpha y_1) F(\alpha
  y_2) F(\alpha y_3) \int^1_0 dx \, \displaystyle{ x \over \left(
      1+x(e^t-1)\right)^\alpha }~.
\end{equation}
The $x$-integral is elementary, and gives:
\begin{equation}
\label{eq:xint}
G(t)=\int d\alpha \, \alpha \, \int [dy] \displaystyle{ F(\alpha y_1)
  F(\alpha y_2) F(\alpha y_3) \over (1-e^{-t})^2 } \left(
  \displaystyle{1\over 2-\alpha} \left( e^{-\alpha t}-e^{-2t} \right)-
  \displaystyle{1\over 1-\alpha} \left( e^{-(\alpha+1)t}-e^{-2t}
  \right) \right)~.
\end{equation}
This is essentially the massive version of Eq.~\ref{eq:lapint} for
$F(\alpha)$, which is singular at $\alpha=2$.  But there is no
singularity in the integrand of Eq.~\ref{eq:xint} for any positive
$\alpha$.  We now see the qualitative (even though quantitatively
inaccurate) solution to the IR renormalon problem, expressed as the
cancellation by what amounts to condensate terms of poles in the
Laplace transform.  It is evident that the same sort of cancellation
takes place in the $\beta$-function, replacing
$e^{-\alpha/bg^2}-e^{-2/bg^2}$ (and, given Eq.~\ref{eq:xint}, another
term is added involving a cancelled pole at $\alpha=1$).

Note, by the way, that it is elementary to derive the differential
equation Eq.~\ref{eq:secdiff} from Eq.~\ref{eq:xint}: Just form
\begin{equation}
{d\over dt}\left(
\displaystyle{e^{-t}\over e^t-1}
\displaystyle{d\over dt}\left((e^t-1)^2 G\right)
\right)~.
\end{equation}
One can look for solutions in the IR or UV by the same method as for
the massless equation.  There are always well-behaved power-series
solutions in $t$, with $G(0)$ as an undetermined parameter; these are
useful in the IR.  In the UV, there is the same solution in powers of
$t^{-1/2}$ (modulo logarithms) as for the massless case
Eq.~\ref{eq:sol1}, but with exponential corrections.  There is also an
exponentially-vanishing solution based on treating the $-G^3$ term as a
perturbation; the leading term is $(e^t-1)^{-2}$ (cf. the massless case
Eq.~\ref{eq:sol2}).  But it appears from numerical studies that this
exponentially vanishing solution develops a singularity for some $t\geq
0$, which we exclude. Thus the only possibility is the usual UV
solution of Eq.~\ref{eq:sol1}, at least for generic values of the
vertex momenta $k_i$.

It might happen that $\phi^3_6$ has a bound state with the quantum
numbers of the $\phi$-field, for some particular value of $k_1^2\not=
0$.  In that case, the Schwinger-Dyson equation for the bound-state
wave function might decrease exponentially in the UV; we have not
studied this possibility in detail.

\subsection{\bf  Numerical Solutions.}

We have used numerical techniques to solve the integral equations
associated with the three models A-C discussed above.  Of these only
model C gives rise to a two-dimensional integral equation; the others
can be put in the one-dimensional form
\begin{equation}
\label{eq:inteq1}
G(t)=\int^\infty_0 dt^\prime K(t,t^\prime) G^3(t^\prime)
\end{equation}
with different kernels $K$. To solve Eq.~\ref{eq:inteq1}, we put it in
the form
\begin{equation}
  G(t) = a+\int^\infty_0 dt^\prime~\tilde K(t,t^\prime) G^3(t^\prime)
\end{equation}
where $\tilde K = K-1/2$ and we treat $a$ as a free parameter, as in
Eq.~\ref{eq:aeq}. For all kernels, $\tilde K$ falls off rapidly when
$t\gg t^\prime$ and one shows $a={1\over 2} \int^\infty_0 dt~G^3$, as
before.  The two dimensional integral equation associated with model C
is treated in a completely analogous manner, with a similar definition
of $a$. We have explored thoroughly the range $0<a\le 2$, and we will
show results for $a=1,2$.  All results were calculated to 16 digits
precision. For $a>2$ it takes progressively longer for our numerical
routines to converge so we cannot exclude the possibility of an upper
bound to $a$ above which solutions cease to exist.

Fig. 8 compares models A $(m=0)$ and B $(m=M).$
For model B $G(0)=a$ (see Eq.~\ref{eq:bc})
which is larger than $G(0)$ for model A,
increasingly so as $a$ gets larger.  This comparison shows the effects
of varying mass assignments.

Fig. 9 compares models B and C.  which have all masses set to the
same non-zero value but differ in the arguments of the $G$'s in the
Schwinger-Dyson equation.  For the same $a$ value (1 in this case) the
differences are remarkably small throughout the entire range of
momenta.  Results such as these give us confidence that toy models,
such as model A and B which can be studied analytically to some
extent are likely to be representative of the full Schwinger-Dyson
equation, even at zero momenta.

The next question to ask is what happens when one goes beyond the cubic
term in the Schwinger-Dyson equation.

\section{\bf AN APPROACH TO SCHWINGER-DYSON VERTEX GRAPHS OF ALL ORDERS}

In this section we take the first incomplete steps towards an analysis
of graphs of all orders in the Schwinger-Dyson equation for $\bar g$
or for the proper vertex $\Gamma$.  These Schwinger-Dyson
graphs are expressed by writing all perturbative vertex graphs which
have no vertex or self-energy insertions; a few such graphs are shown
in Fig. 6.  Then in the full Schwinger-Dyson equation each such graph
is replaced by one of the same topology but with full vertices and
propagators.  In what follows we do not need to consider the cubic
graph which was the concern of previous sections.

All remaining vertex skeleton graphs have the property that the minimum
number of lines in any closed loop is at least four; graph theorists
say that such graphs have girth four.  It turns out that such bare
skeleton graphs, in perturbative $\phi^3_6$, have exactly one UV
logarithm ($\ln\Lambda_{\rm UV}^2$, where $\Lambda_{\rm UV}$ is a UV
cutoff) and no IR logarithm ($\ln M^2$).  This follows
straightforwardly from an elementary remark about $\phi^4_4$ graphs, as
we now show.

Every perturbative skeleton graph of $N$ loops, ($3N$ lines, $2N+1$
vertices) has the following Feynman-parameter form in Minkowski space,
which is for the moment convenient:
\begin{equation}
\label{eq:par}
\Gamma^0_N={\rm constant}\times~g^{2N} \int {[dx]\over U^3}
\ln\biggl({-\phi/U+\sum x_iM_i^2-i\epsilon\over \Lambda_{\rm
    UV}^2}\biggr)
\end{equation}
where $[dx] = \delta(1-\sum x_i) \prod^{3N}_{i=1} dx_i$.  $U$ is the
graph's determinant (a positive sum of monomials of order $N$ in the
$x_i$), and $\phi$ (not to be confused with the field $\phi$) is a sum
of positive polynomials in the $x_i$ times scalar momentum functions
$k_a\cdot k_b$.  The single UV divergence is explicit, and can be
removed by taking the imaginary part of $\Gamma_N^0$, replacing the
logarithm by a $\theta$-function.  Any other divergences in the
imaginary part must come from the vanishing of $U$.  But, as is well
known, $U$ can vanish only when at least all the parameters of a loop
in the graph go to zero.  Let $\{x_i\}$ be the parameters of one and
only one such loop of $n$ lines, and replace $x_i$ by $\lambda x_i$,
$\lambda\rightarrow 0$.  Then $U$ vanishes like $\lambda$.  But $[dx]$
behaves like $\lambda^{n-1}d\lambda$ times irrelevant factors, and the
integral is finite if $n\geq 4$, that is, for girth-four graphs.  If
all the parameters of two complete (possibly overlapping loops) are
scaled with parameters $\lambda_1,\lambda_2$, then $U$ vanishes like
$\lambda_1\lambda_2$, and the argument can be repeated, with some
necessary variations.  The reader may easily check that only for
skeleton graphs the total number of distinct lines in $\ell$ complete
loops is at least $3\ell+1$ for any skeleton graph of order $N>\ell$.
Then the one-loop scaling argument given above can be repeated, and one
shows there are no singularities in the Feynman parameter integral.
(The case $N=\ell$ is trivial, since not all the $x_i$ can be scaled to
zero.)

The Minkowski-space imaginary part of $\Gamma_N^0$ in Eq.~\ref{eq:par}
is, when all masses are equal to $M$,
\begin{equation}
  {\rm Im}\Gamma^0_N={\rm constant}\times~g^{2N}\int {[dx]\over U^3}~
  \theta\biggl({\phi\over U}-M^2\biggr)~.
\end{equation}
Consider now the kinematics of Fig.~1 where $k_1,-k_2$, and $-k_3$ are
forward timelike, with $k_1^2=s>4M^2,~k_2^2=k_3^2=M^2$.  One easily
shows $k_1\cdot k_2=k_1\cdot k_3=s/2,~k_2\cdot k_3= s/2-M^2$, which is
enough to show that the argument of the $\theta$-function is positive
(trivially so at $M=0$) and thus the $\theta$-function can be replaced
by unity.  We then reduce the problem of finding ${\rm Im}\Gamma_N^0$
to summing the integrals
\begin{equation}
  \int {[dx]\over U_i^3}
\end{equation}
where $i$ runs over all the $N-{\rm th}$ order skeleton graphs.  This
problem (with $U_i^{-3}\rightarrow U^{-2}_i$) has been addressed for
$\phi^4_4$\cite{cornmor}, using the following steps which we will
repeat here.

1.  Count all the $N^{\rm th}$ order graphs, and note that the number
of these which are not skeleton graphs vanishes at large $N$ relative
to the number of skeleton graphs.  We denote this number as $Q_N$.

2.  Find, by a combination of analysis and numerical work, the number
$C_i$ of monomials in $U_i$.  This number is called the complexity of
the graph, and is equal to the number of spanning trees.

3.  Show that
\begin{equation}
  \langle U\rangle_i \equiv {\int [dx]U_i\over \int [dx]}=
  C_{i}{(3N-1)!\over (4N-1)!}~.
\end{equation}
That is, every monomial in $U_i$ contributes exactly the same to
$\langle U \rangle_i$.

 4.  Use a H\"older inequality to show
\begin{equation}
  \biggl\langle {1\over U} \biggr\rangle_i^3 \geq {1\over \langle
    U\rangle^3_i}~.
\end{equation}

5.  Observe that the distribution of $C_i$ for all the $N$-loop graphs
centers around an average value $\langle C \rangle.$

We then have, inserting the usual powers of $2\pi$,
\begin{equation}
{\rm Im}\Gamma_N^o \gwig \biggl({g^2\over
(4\pi)^3}\biggr)^N Q_N~
\displaystyle{1\over \langle U \rangle^3_i}~.
\end{equation}

Now Bender and Canfield\cite{bender} have shown, in a deep
graph-theoretic result, that
\begin{equation}
\label{eq:bencan}
Q_N \buildrel \cdot \over = \biggl({3\over 2}\biggr)^N~N!
\end{equation}
and we have made a numerical analysis of large-$N$ graphs to find
\begin{equation}
  \langle C \rangle \buildrel \cdot \over = \biggl({16\over
    3}\biggr)^N~.
\end{equation}
(Here $A_N \buildrel \cdot \over = B_N$ if $A_N$, $B_N$ are each of the
form $c_o(N!)^{c_1} {c_2}^N N^{c_3} \bigl(1+O(1/N)\bigr)$ at large $N$,
and the constants $c_1$ and $c_2$ are the same for $A_N$ and $B_N$.)

Putting everything together,
\begin{equation}
\label{eq:410}
{\rm Im}\Gamma_N^o \buildrel \cdot \over \geq N!
\biggl[{a~g^2\over (4\pi)^3}\biggr]^N~, \quad
a={2^{11}\over 3^8} \simeq 0.312~.
\end{equation}
As expected for an asymptotically free theory, this is not
Borel-summable.

Even though extracted from a high-energy process in Minkowski space,
the result of Eq.~\ref{eq:410} looks exactly like what one would expect
from a lowest-order (semi-classical) instanton calculation which
requires zero external momenta and masses.  There is no sign of
external momenta or masses in Eq.~\ref{eq:410}, which renders this
formula compatible with the Lipatov procedure.  So let us compare the
coefficient $a$ in Eq.~\ref{eq:410} to the instanton result.  As is
well known\cite{houghton}, there are instantons in massless $\phi^3_6$,
of the form
\begin{equation}
\phi_{c\ell}={48\rho^2\over
g\bigl[(x-a)^2+\rho^2\bigr]^2}
\end{equation}
whose action is
\begin{equation}
I=\biggl({12\over 5}\biggr){(4\pi)^3\over g^2}~.
\end{equation}
This means that using the Lipatov procedure one would find
an expression of the form of
Eq.~\ref{eq:410} for ${\rm Im}\Gamma_N^o$, with $a=5/12=0.417$.  Our
lower-limit value $2^{11}/3^8$ is about 3/4 of this (presumably
correct) value, and further numerical work, which we have not done,
would bring this lower-limit value closer to the true value of $a$.

The fact that there are no logarithms in ${\rm Im}\Gamma_N^o$ (or only
one in $\Gamma_N^o$) is of vital importance for analyzing the
Schwinger-Dyson equation, when full propagators and vertices are used
to dress the bare skeleton graph.  This amounts to inserting $2N+1$
factors of $\bar g$ (see Eq.~\ref{eq:running}), where every $\bar g$
behaves like $(b~\ln k^2)^{-1/2}$ for the appropriate momentum $k$.
One should then not be surprised to find that the $N$-loop full
skeleton graphs behave, for large external momentum $k$, like $(\ln
k^2)^{-{1\over 2}(2N+1)+1}$ and that the sum of full skeleton graphs is
of the form Eq.~\ref{eq:410}, with $g^2$ replaced by the (one-loop)
running charge $\bar g^2$.  The Schwinger-Dyson equation thus yields
the results of RG-improved instanton calculations, with no explicit
introduction of instantons.

We will not show the derivation of this result in detail, but merely
sketch it.  As in Section~III we make the approximation that every
$\bar g$ depends only on $\sum k_i^2$, where $k_i$ are the vertex
momenta, and introduce the Laplace transform
\begin{equation}
\bar g(k_i)=\int d\alpha F(\alpha)~e^{-\alpha~\ln(\sum k_i^2)}~.
\end{equation}
Anticipating that the leading UV behavior of the $N^{\rm th}$
order graphs is dominated by the UV behavior of the cubic graph
studied in Section 3, we choose
\begin{equation}
\label{eq:fchoose}
F(\alpha)=(\pi b\alpha)^{-1/2}
\end{equation}
corresponding to $\bar g \rightarrow (b\ln k^2)^{-1/2}$.

As before, the factor $\exp(-\alpha \ln \sum k_i^2)$ is written as
$(\sum k_i^2)^{-\alpha}$ and the momentum integrals are done after
Feynman parameterization.  A tedious analysis, not given here, shows
that one need not introduce $2N+1$ new Feynman parameters for the new
denominators $\sum k_i^2$; instead, one shows that the correct
asymptotic behavior is also found by replacing such denominators by
those of neighboring propagators.  The result is that many of the
propagators in a bare skeleton graph appear to powers $-(1+\alpha_i)$,
where $\alpha_i$ is the Laplace transform variable of a neighboring
vertex.  Finally one finds, setting the mass $M$ to zero, and going to
asymptotically-large $k$ as in Section 3,
\begin{equation}
\bar g_N = Q_N\biggl[{1\over \pi(4\pi)^3b}\biggr]^N \int
\Pi \alpha_i^{-1/2}d\alpha_i \Gamma(\sum \alpha_i)
\bigl[\Pi\Gamma(1+\alpha_i)\bigr]^{-1} \int
{[dx]\over U^3} \Pi x_i^{\alpha_i}
\biggl({\phi\over k^2 U}\biggr)^{-\sum\alpha_i}
\end{equation}
where $\phi$ and $U$ are the appropriate functions for the bare
skeleton graph, and an irrelevant constant factor has been dropped.  In
the result of Eq.~\ref{eq:fchoose}, $Q_N$ is the number of $N^{\rm
  th}$-order skeleton graphs, given in Eq.~\ref{eq:bencan}.

In the kinematics of Fig.~7, $\phi$ is of the form of $k^2$ times a
function of $y_i$.  We scale the $\alpha_i$ by $\alpha_i=\alpha y_i$,
with $\sum y_i=1$, and get
\begin{equation}
\label{eq:gbar}
\bar g_N=\int^\infty_0 d\alpha {\alpha^{N-{3\over 2}}\over
\Gamma\left(N-{1\over 2}\right)} e^{-\alpha\ln k^2} H(\alpha)
\end{equation}
where
\begin{equation}
\label{eq:bigh}
H(\alpha) = \Gamma(N-{1\over 2})
\biggl[{1\over \pi(4\pi)^3b}\biggr]^N Q_N \int [dy]
\Pi y_i^{-1/2} \Gamma(1+\alpha)
\int {[dx]\over U^3} \Pi x_i^{\alpha_i}
\biggl({\phi\over k^2U}\biggr)^{-\alpha}~.
\end{equation}
In Eq.~\ref{eq:gbar} and Eq.~\ref{eq:bigh} we wrote $\Gamma(\sum
\alpha_i)\equiv \Gamma(\alpha)$ as $\alpha^{-1}\Gamma(1+\alpha)$ and
removed a factor $\Gamma(N-{1\over 2})$ for simplicity.  This allows us
to assert that $H(\alpha)$ is regular at $\alpha=0$, and in fact one
checks easily that
\begin{equation}
H(0) \buildrel \cdot \over =
\biggl[{1\over \pi(4\pi)^3b}\biggr]^N
Q_N \int {[dx]\over U^3}
\end{equation}
which is just the function appearing in the sum of bare skeleton
graphs.  It is clear from Eq.~\ref{eq:gbar} that the leading asymptotic
behavior of $\bar g_N$ comes from $\alpha\simeq 0$, which gives
\begin{equation}
\label{eq:gbar2}
\bar g_N \buildrel \cdot \over = H(0) \int d\alpha {\alpha^{N-{3\over
      2}}\over \Gamma(N-{1\over 2})} e^{-\alpha\ln k^2} \buildrel \cdot
\over = N!\biggl[{a\over (4\pi)^3 b\ln k^2}\biggr]^N
\end{equation}
using the previous result for $H(0)$.

This is as far as we will carry the analysis. We have no particularly
good ideas for extending the toy models of Section~III to all orders in
a controllable way.  Nor will we attempt to find some physical way to
sum the series of Eq.~\ref{eq:gbar2} (based, e.g., on decay of the
meta-stable vacuum).  If the sum of terms is represented by a (formal)
Borel transform, there arises an ambiguity of the form $(k^2)^{-c}$,
where $c=(4\pi)^3 b/a$.  This is of the usual type associated with
RG-improved instantons.

\section{\bf CONCLUSIONS}

In the long run, perturbative QCD must fail because of its IR
singularities, and some sort of new perturbative analysis of
Schwinger-Dyson equations will have to be seriously attempted.  This is
much more difficult than $\phi^3_6$, because of the complications of
spin, gauge dependence, and confinement.  Still, we believe that some
of our results will hold for QCD, in particular, the cancellation of
poles in Borel transforms as in Eq.~\ref{eq:regborel}. Currently there
is considerable phenomenological interest\cite{zak} in renormalon
problems, and it would be valuable to test the mechanism we identify as
relevant, versus other prescriptions such as principal part
integration.

A similar mechanism may well hold for Borel transforms associated with
instantons, although we are unable to say in $\phi^3_6$ because of its
instability.  In QCD (or a similarly-stable asymptotically free theory,
like the $d=2$ nonlinear $\sigma$ model), it is quite important to
learn how to deal with phenomena canonically associated with instantons
(e.g., the Lipatov technique) by other means than the usual
RG-corrected semi-classical results, because the perturbative
corrections to semi-classical instanton physics involve IR
singularities such as renormalons.  Understanding such phenomena
directly from the Schwinger-Dyson equations is a worthwhile thing to
do.  Moreover, at the moment Schwinger-Dyson equations are the most
straightforward way of dealing\cite{cornmor} with the
non-Borel-summable divergences at large Minkowski-space external
momenta.  We hope to report in the future on future progress in this
direction, going well beyond the first steps of Section~IV.

We have seen that all the toy models of the Schwinger-Dyson equation
have solutions with a free parameter, equivalent to the
renormalization mass $\mu$.  Since $\mu$ comes in as a tool in dealing
with amplitudes needing regularization, and since asymptotically free
Schwinger-Dyson equations in their canonical form need no such
regularization, this is a bit of a surprise.  On the other hand, one
would be surprised if a single Schwinger-Dyson equation---in our case,
that for the running charge defined in Eq.~\ref{eq:running}---had a
single solution for the three-point function, making no reference to
higher-point functions as the usual hierarchy argument specifies.  It
remains to be seen how this is settled by consideration of further
Schwinger-Dyson equations.

\section{\bf ACKNOWLEDGMENTS}

One of us (JMC) was supported in part by NSF Grant PHY-9218990.  Both
of us thank J. Ralston and W. Newman for useful conversations.

\newpage
\section*{FIGURE CAPTIONS}
\begin{enumerate}
\item Bethe-Salpeter equation for the renormalized three-point function
  (small blobs) expressed in terms of the two-particle-irreducible (in
  the $k_1$ channel) four-point function (large blob) and the vertex
  renormalization constant $Z_1.$ All propagators are fully dressed.
\item Bethe-Salpeter equation for the renormalized three-point function
  (small blobs) expressed in terms of the one-particle-irreducible (in
  the $k_1$ channel) four-point function (large blob) and the vertex
  renormalization constant $Z_1$. All propagators are fully dressed.
\item Bethe-Salpeter equation for the renormalized inverse propagator
  in terms of the renormalized three-point-function and the vertex
  renormalization constant $Z_1$. All propagators are fully dressed.
\item Lowest order approximations to the Schwinger-Dyson equations for
  a) the three point function and b) the inverse propagator.
\item Cubic Schwinger-Dyson equation in $g|\psi|^2\phi$ theory.
\item Schwinger-Dyson equation for the running charge (solid blob).
  All propagator lines are free propagators with renormalized masses.
\item Momentum assignments for the truncated Schwinger-Dyson equation
  for the running charge.
\item Numerical solutions of the toy-model Schwinger-Dyson equation for
 models A and B for $a=1,2$ where $a = 1/2 \int \, dt' G^3(t').$
\item Numerical solutions of the toy-model Schwinger-Dyson equation for
  models B and C for $a=1.$
\end{enumerate}
\end{document}